\documentclass[aps,preprint]{revtex4}

\pdfoutput=1

\usepackage[utf8]{inputenc}
\usepackage{lmodern}
\usepackage{hyperref}
\usepackage{amsmath} 
\usepackage{graphicx}
\usepackage{amssymb}
\usepackage{amsthm}
\usepackage{epstopdf} 
\usepackage{natbib}
\usepackage{dcolumn}
\usepackage{amsfonts}

\newcommand{\si}{}
\newcommand{\micro}{$\mu$}

\newcommand{\cc}{\,\mathrm{cm^{-3}}}
\newcommand{\wcm}{\,\mathrm{W/cm}^2}
\newcommand{\mic}{\,\mu\mathrm{m}}

\begin{document}

\title{Effect of the laser wavefront in a laser-plasma accelerator}

\author{B. Beaurepaire, A. Vernier, M. Bocoum, F. B\"ohle, A. Jullien, J-P. Rousseau, T. Lefrou, D. Douillet, G. Iaquaniello, R. Lopez-Martens, A. Lifschitz and J. Faure}

\affiliation{Laboratoire d'Optique Appliqu\'ee, UMR 7639 ENSTA-CNRS-Ecole Polytechnique, 91761 Palaiseau, France}%

\begin{abstract}
A high repetition rate electron source was generated by tightly focusing kHz, few-mJ laser pulses into an underdense plasma. This high intensity laser-plasma interaction led to stable electron beams over several hours but with strikingly complex transverse distributions even for good quality laser focal spots. Analysis of the experimental data, along with results of PIC simulations demonstrate the role of the laser wavefront on the acceleration of electrons. Distortions of the laser wavefront cause spatial inhomogeneities in the out-of-focus laser distribution and consequently, the laser pulse drives an inhomogenous transverse wakefield whose focusing/defocusing properties affect the electron distribution. These findings explain the experimental results and suggest the possibility of controlling the electron spatial distribution in laser-plasma accelerators by tailoring the laser wavefront. 

\end{abstract}

\maketitle

Laser-plasma accelerators \cite{esar96,esar09} currently provide 100 MeV to few GeV electron beams in short distances \cite{faur06,leem06}, owing to their very high accelerating gradients $\simeq 100$ GV/m. In these experiments, an intense laser pulse drives a large amplitude plasma wave in which plasma electrons can be trapped and accelerated to relativistic energies in millimeter distances. These laser driven electron sources are also of great interest because the generated electron bunches can be extremely short, with durations reaching down to a few fs only \cite{lund11}. In general, laser-plasma accelerators operate using 100 TW, Joule level laser systems. In this case, the laser power exceeds the critical power for relativistic self-focusing \cite{sun87,bori92}, $P/P_c>1$, and the laser pulses are self-focused over several Rayleigh lengths. Injection and acceleration of electrons occur in the region where the laser is self-focused, corresponding to the region where the plasma wave amplitude is high. In many experiments, the laser wavefront is corrected in order to obtain the best laser spot in the focal plane and little attention is given to the laser transverse distribution outside the focal plane. 

In contrast, several groups have recently been developing laser-plasma accelerators operating at high repetition rate, using kHz lasers with energies $< 10$ mJ. Such developments are particularly important for applications as the high repetition rate improves the beam stability and permits data accumulation. Indeed, kHz electron beams with femtosecond duration and energies in the MeV range are of great interest for ultrafast electron diffraction \cite{scia11,musumeci10}, a powerful technique for investigating structural dynamics in matter with femtosecond resolution. Recent experiments have shown the possibility of generating 100 keV electron bunches at kHz repetition rate \cite{he13}. Electron bunches from a kHz laser-plasma source were successfully used to obtain clear diffraction patterns on single crystal gold foils \cite{he13b}. In these high repetition rate experiments, the laser pulse has to be focused very tightly in order to reach intensities in excess of $10^{18}\wcm$, as required to drive large amplitude plasma waves. Therefore, the Rayleigh range of the laser beam is extremely short, typically on the order of $z_R\simeq 10\mic$. In addition, self-focusing does not occur because the power is not high enough, $P/P_c<1$. Consequently, the interaction region extends outside the focal region and one expects that experimental results should also depend critically on the laser distribution in the intermediate field. 

In this letter, we show results from a laser-plasma accelerator operating at kHz repetition rate. The interaction region was large compared to the Rayleigh range, $L\gg z_R$ so that the laser distribution outside the focal region plays an important role in the physics of electron injection and acceleration. We demonstrate that the spatial distribution of the electron beam originates from the inhomogeneity of the laser distribution outside the focal region, due to an imperfect laser wavefront. PIC simulations show that an inhomogeneous laser pulse drives an inhomogeneous transverse wakefield which affects the electron beam quality through its focusing/defocusing properties. Our work emphasizes the influence of the laser wavefront over wake excitation as well as on the electron beam spatial distribution.

The experiment was performed using $1\,$kHz, $800\,$nm laser pulses with $\tau=22\,\mathrm{fs}$ duration at Full Width Half Maximum (FWHM). The energy on target was in the range $2.5-3\,\mathrm{mJ}$. The laser system provides high contrast laser pulses and good beam quality through the use of spatial filtering in a hollow core fiber \cite{jull14}. The laser pulses were focused down to a spot size of $\lesssim2\,\mu\mathrm{m}$ FWHM using a $f/1.5$, $90^\circ$ off-axis parabolic mirror, thus leading to a peak intensity of $I = 2.5-3\times10^{18}$\,W/cm$^{-2}$ and a measured Rayleigh range of $\simeq 20\,$\si{\micro}m. The gas target consisted of a $100\mic$ diameter gas jet operating in continuous flow so as to make kHz operation possible. The electron beam profile was measured using a CsI(Tl) phosphor screen imaged onto a 14-bit CCD camera. The electron energy distribution could be obtained by inserting a magnetic electron spectrometer. Information on laser propagation in the gas jet was obtained by imaging the transmitted laser mode at the exit of the plasma. Finally, the electron density was monitored \textit{in situ} by transverse interferometry with a 22$\,$fs probe laser pulse. The experiment was run at kHz repetition rate and all data presented in this paper is averaged over 500 to 1000 shots.
 
 \begin{figure}[t!]
\centerline{\includegraphics[width=9cm]{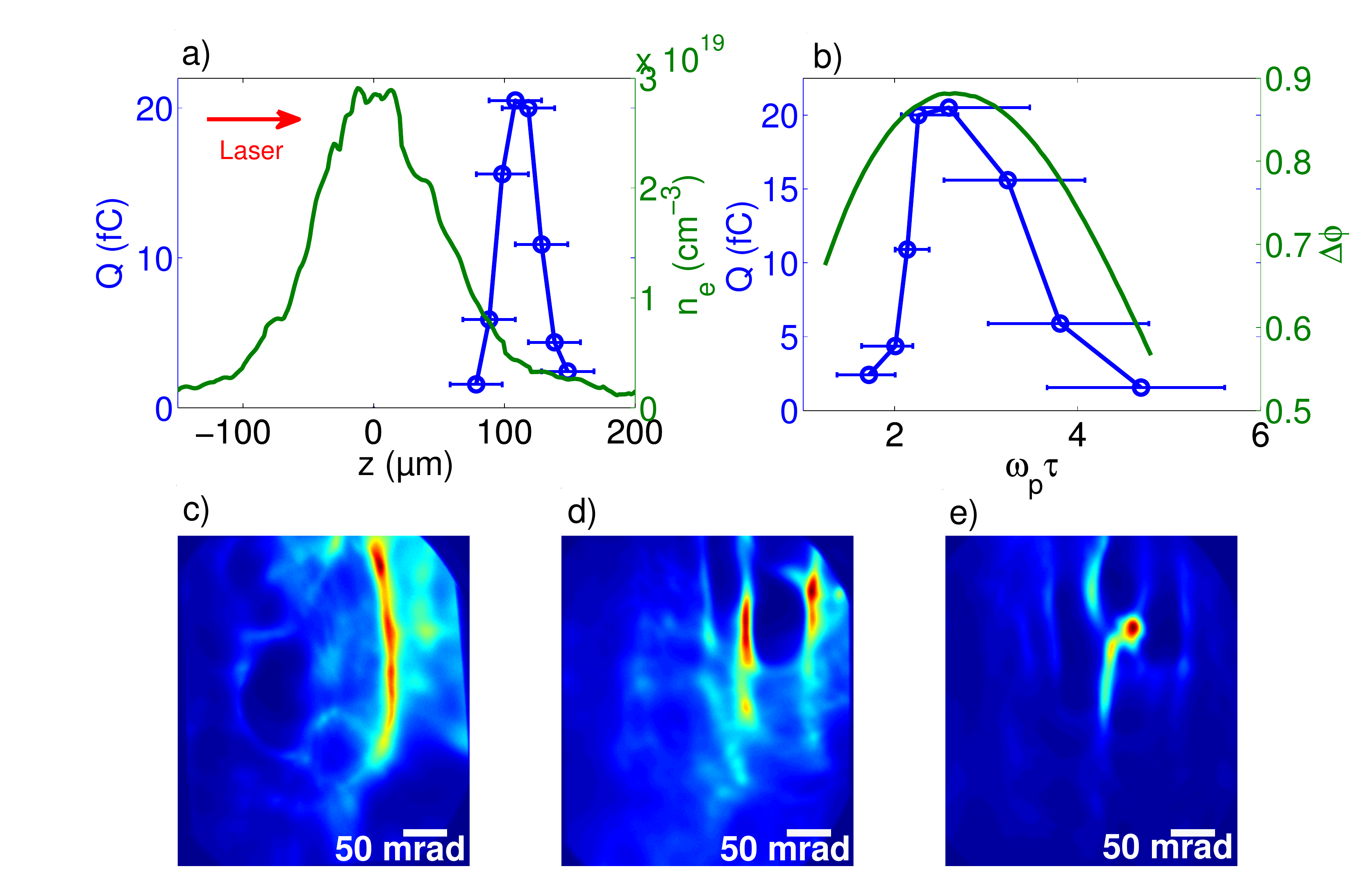}}
\caption{(a) Green curve: measured longitudinal electron density profile. Blue curve: charge/shot of the accelerated electrons for different positions of the laser focus. (b) Beam charge as a function of $\omega_p(z)\tau$ where $\omega_p(z)$ is estimated using the experimental plasma density at the focal plane. The green curve is the theoretical plasma wave amplitude $\Delta \phi$. Horizontal error bars originate from a $\pm20\mic$ uncertainty on the position of the focus. (c) (d) and (e) are typical images of electron beam profiles.}\label{fig1}
\end{figure}

Experiments were mostly performed in Nitrogen gas. The green curve in fig.~\ref{fig1}.a) represents a typical longitudinal electron density profile when the laser is focused $150\mic$ above the gas jet. Here the density reaches $n_e=3\times 10^{19}\cc$ and the density gradients are on the order of $50\mic$. By measuring the transmitted laser mode, we have checked that the laser pulse does not undergo filamentation and that ionization induced defocusing \cite{rank91,ches99} does not significantly affect laser propagation (see Suppl. Mat.). Therefore, it is safe to consider that the laser pulse can propagate over a distance $L\gg z_R$ in the ionizing Nitrogen gas without significant distortion before it reaches the focus.


The blue curve in fig. \ref{fig1}.a) represents the measured accelerated charge as a function of the position of the laser focus. We found that the amount of
accelerated electrons depends critically on the position of the focal plane, and that a charge as high as 20 fC/shot (i.e. 20 pC/s) can be obtained when the laser is focused in the density downramp. In fig. \ref{fig1}.b), the accelerated charge is plotted as a function of $\omega_p(z)\tau$, where $\omega_p(z) =\sqrt{\frac{n_e(z)e^2}{m_e\epsilon_0}}$ is the local plasma frequency along the propagation axis. The green curve represents the plasma wave amplitude $\Delta\phi$ obtained from a 1D nonlinear fluid model of plasma wave excitation \cite{teyc93}. The experimental data peaks at $\omega_p\tau \approx 2.6$ which is very close the theoretical estimate of resonant plasma wave excitation, i.e. for a 22 fs laser pulse, the resonant density is $n_e^{res}\simeq 5\times10^{18}\cc$. This data indicates that electrons are generated when high amplitude plasma waves are resonantly excited by the laser pulse. In addition, fig.~\ref{fig1}.a) suggests that electrons are injected in the wakefield by density gradient injection \cite{bula98,bran08,gedd08,faur10,schm10}, in a regime similar to Ref.~\cite{he13}. Comparable results were obtained using Helium instead of Nitrogen, excluding ionization injection \cite{mcgu10,pak10} as a possible injection mechanism. Finally, we measured that the accelerated electrons have a large energy distribution extending to about $100\,$keV (see Suppl. Mat.). 

Figures \ref{fig1}.c)-e) show typical electron beam transverse distributions obtained during different experimental runs. Each image was obtained by averaging over about 1000 shots. These beams present complex structures which are not random and are remarkably stable over the course of several hours. In these kHz experiments, the electron beam parameters vary very little for a given set of experimental conditions: the charge fluctuations are on the order of $7\%$ (see Ref.~\cite{he13b} and Suppl. Mat.) and the energy and divergence fluctuate at the percent level. The shape of the beam does, however, change from run to run. Depending on the experimental conditions, it can exhibit various complex patterns (see fig.~\ref{fig1}.c)-d)) or a low-divergence beam  ($<10\,$mrad and containing $\simeq20\,$fC). Such electron distributions are quite different from those emerging from high-energy laser-plasma accelerators. Similar beams were obtained in Ref.~\cite{he13}, but no explanation was given for the observed complex structures.

\begin{figure}[th]
\centerline{\includegraphics[width=9cm]{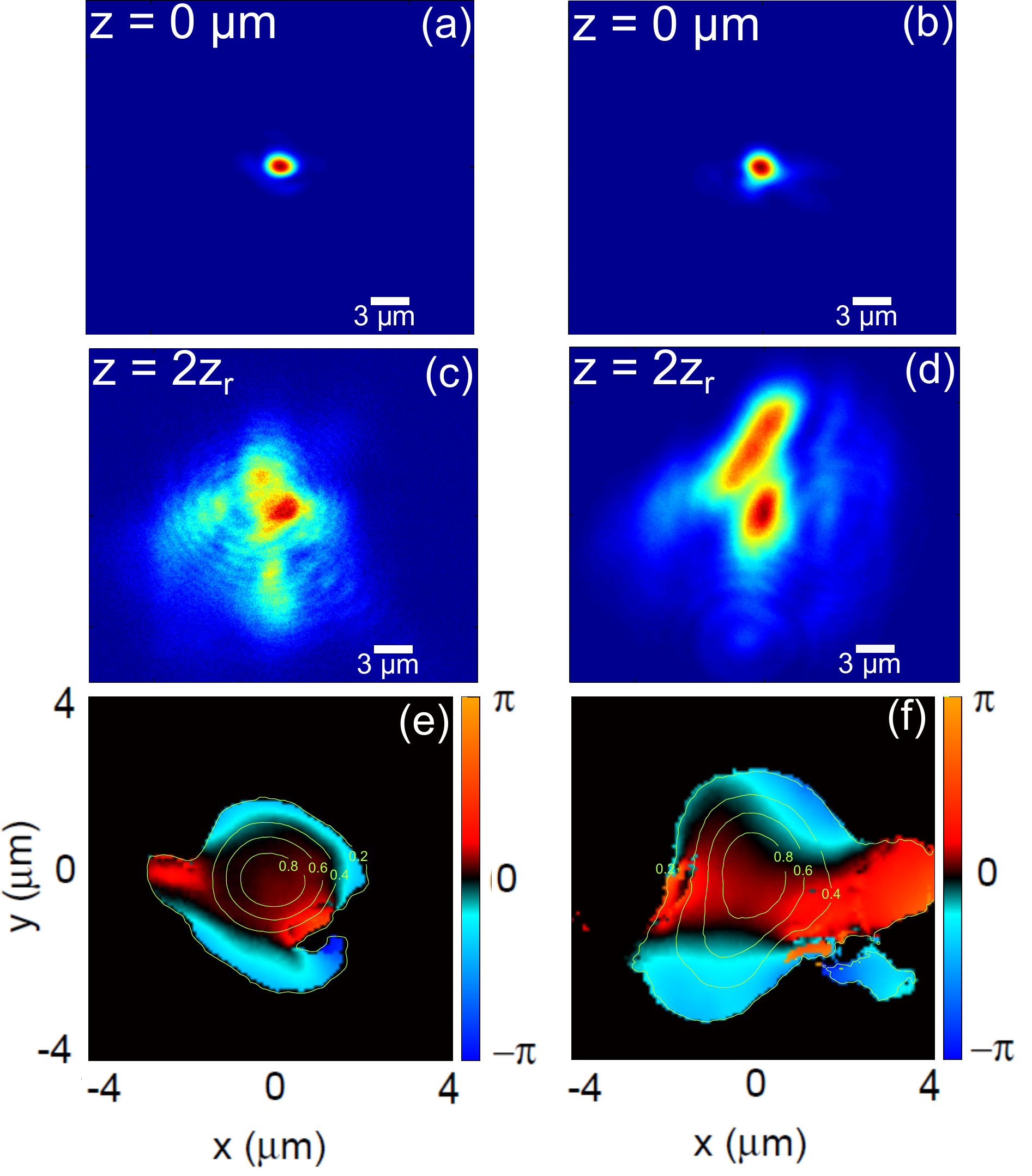}}
\caption{Laser focal spots and wavefronts. a) and b) are the laser focal spots for two different runs. c) and d) are the corresponding out-of-focus laser transverse distribution at $z=2z_R$. e) and f) are the corresponding reconstructed laser phase at the focal plane.}\label{fig2}
\end{figure}

The laser focal spot does not change significantly in the various
experimental runs, however, we noticed that the laser distribution in
the intermediate field can be quite different from one run to the
next. Figure \ref{fig2}.a) and b) show the laser beam at focus for two
different runs: both beams have a $\sim 2\mic\,$diameter FWHM, and the laser
intensity is similar. In contrast, the differences
between the laser distributions at an intermediate plane ($z=2z_R$, fig. \ref{fig2}.c) are conspicuous. The strong degradation of the beam
quality far from the focal plane indicated the presence of significant wavefront distortions, which can vary from run to run.

The wavefront  was reconstructed numerically from the laser intensity
distribution at the two planes shown in fig. \ref{fig2} ($z=0$ and
$z=2 z_R=40 \mic$) using the Gerchberg-Saxton algorithm
\cite{gerch72}. To test the accuracy of the result, the obtained
phase was used to reconstruct the intensity distribution before the
focal plane  (at $z=-2z_R$). The extrapolated intensity was in good agreement
with the experimentally measured laser distribution in this plane. The wavefronts
obtained at the focal plane are shown in fig. \ref{fig2}.e) and
f). They exhibit complex structures, which are quite different for the two runs. This data suggests that the run-to-run variation of the laser wavefront has a direct influence over the electron beam distribution.

To understand how the wavefront distortions actually affect the electron acceleration, we performed Particle-in-Cell (PIC)
simulations including realistic laser wavefronts. We used the code Calder-Circ \cite{lifs09}, a fully electromagnetic
3D code based on cylindrical coordinates $(r,z)$ and Fourier decomposition in the poloidal direction. The simulations were performed
using a mesh with $\Delta z=0.3\,k_0^{-1}$ and $\Delta r=1.5\,k_0^{-1}$ (where $k_0$ is the laser wave vector),
and the 7 first Fourier modes. The neutral gas density profile was taken from
the experimental data. The simulations start with pure neutral Nitrogen, which
is ionized via tunnel ionization. The number of macro-particles per
cell before ionization is 2000, which corresponds to 2000$\times$5=10000 macro-electrons per cell in the
region of full ionization of the L-shell of Nitrogen. 
We performed simulations on two test cases. In the first one, the laser
intensity was taken from the experimental data and the wavefront
was reconstructed using the Gerchberg-Saxton algorithm. In the second
one, the same experimental laser intensity was used, but the wavefront was assumed to be perfectly flat.

\begin{figure}[t]
\centerline{\includegraphics[width=9cm]{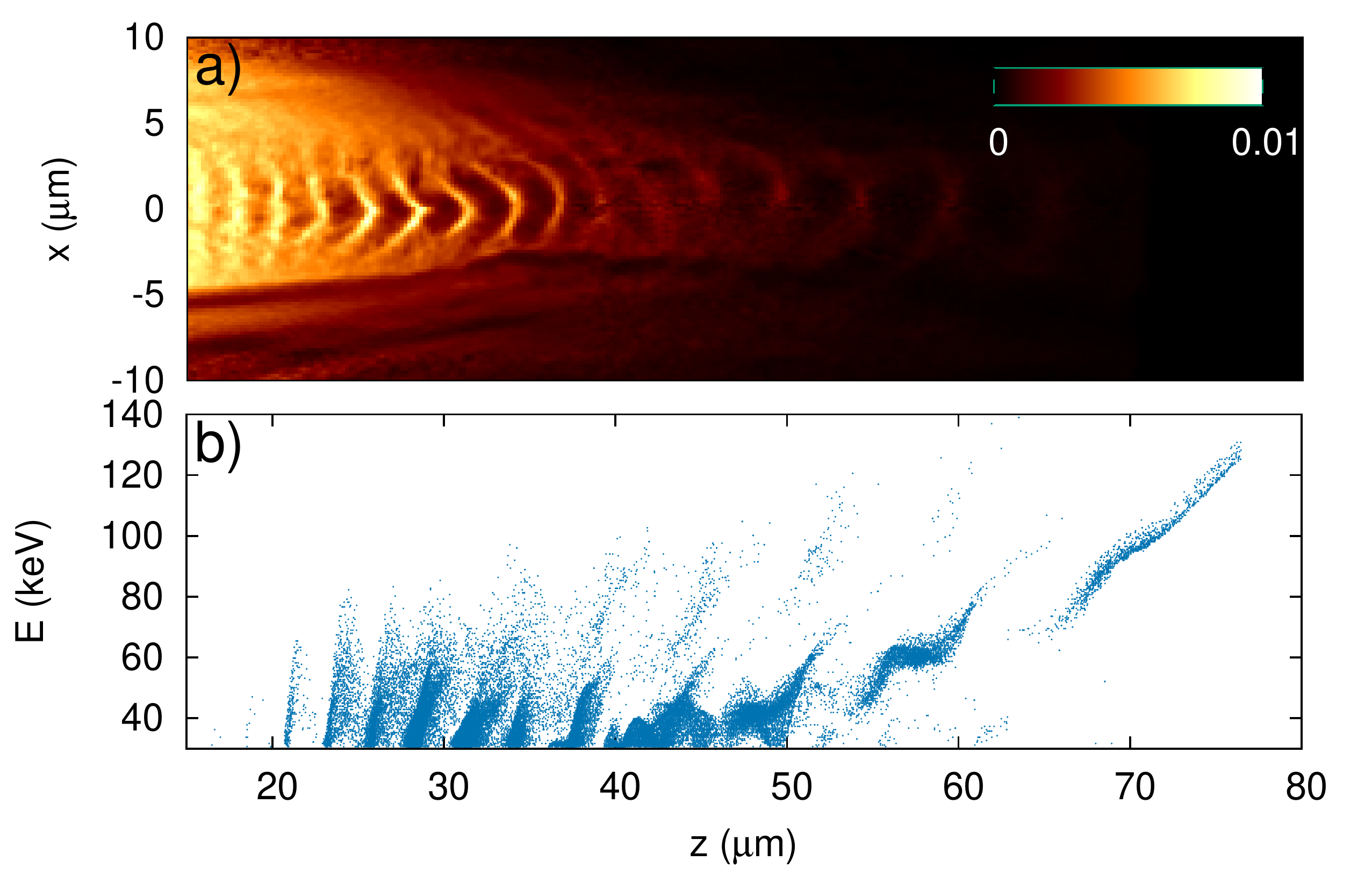}}
\caption{a) Results of 3D PIC simulations with reconstructed laser wavefront. Electron density $n_e/n_c$  in the downward density ramp at the exit of the gas jet, at $t=1560\,\omega_0^{-1}$. b) Corresponding electron longitudinal phase space showing electron trapping and acceleration in the wakefield.}\label{fig3}
\end{figure}

PIC simulations confirm that electrons are injected in the wakefield generated in the downward density gradient. Figure \ref{fig3}.a) shows a map of the electron density: a large amplitude plasma wave is excited by the laser pulse, and traps and accelerates numerous electrons at the 50-100$\,$keV level, fig.~\ref{fig3}.b). Analysis of the simulations shows that the wake phase velocity decreases with time because of the density gradient and this eventually leads to the trapping of electrons \cite{bran08,faur10} (see also Suppl. Mat. for more details). In the simulation, trapping occurs when the phase velocity reaches $v_p/c\simeq 0.45$. This low phase velocity explains why electrons are accelerated in the 100 keV energy range. Note also that trapping occurs after the laser pulse has passed, i.e. the electron bunch sits $\simeq70\mic$ behind the laser pulse and never interacts with the laser field. Finally, the energy distribution at the exit of the plasma agrees very well with the experiment (see Suppl. Mat.) which confirms the trapping and acceleration mechanism.


\begin{figure*}[ht!]
\centerline{\includegraphics[width=14cm]{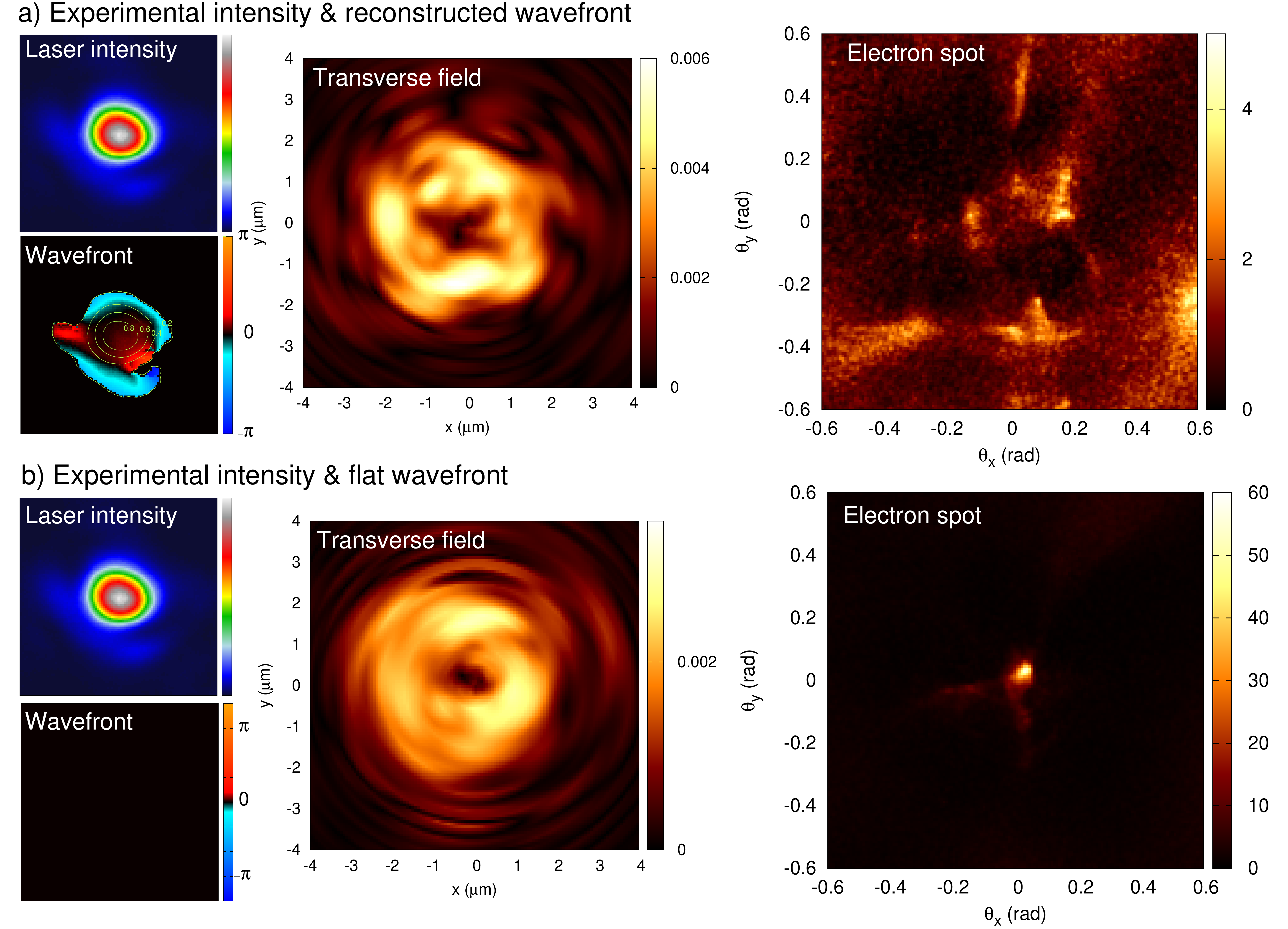}}
\caption{Simulations results for  a laser pulse with
            both  intensity distribution and wavefront taken from
            experimental data (a) and with
            the same experimental intensity distribution and a flat wavefront
            (b). Left column: input intensity distribution
            and wavefront at the focal
          plane. Central column: screenshots with the cross section of the
          transverse field $E_\bot/E_0$ ($E_0=m_ec\omega_p/e$ is the cold wave breaking field) for the time and the region of injection
          ($\tau=1560 ~\omega_0^{-1}$, $z=28~ \mu$m). Right column:
          electron spot far from the gas jet.}\label{fig4}
\end{figure*}

The trends described above are also found in the simulations
using a flat wavefront. However, differences are dramatic
when inspecting the shape of the electron transverse distribution obtained in each
case. We now focus on the impact of the transverse wakefield $E_\bot=\sqrt{E_x^2+E_y^2}$ on the electron distribution. When the experimental laser spot and the reconstructed wavefront are used (fig.~\ref{fig4}.a), the cross sections of the transverse wakefield $E_\bot$ are highly
asymmetric and the electron distribution is complex, with hot spots and
regions without electrons. This electron distribution is similar to the
experimental ones (fig.~\ref{fig1}.c and d). For the case with
a flat wavefront, fig.~\ref{fig4}.b), the cross section of the transverse wakefield is more rotationally symmetric, and as a result, the electron bunch is much more collimated. This flat phase case resembles the experimental data in fig.~\ref{fig1}.e), indicating that on that particular experimental run, it is likely that the wavefront was less distorted. 

The total charge ($\sim 40\,$fC, close to the experimental result) is similar when a flat or distorted wavefront is used, but the crucial difference is that a flat wavefront yields an electron beam contained within a small solid angle $\lesssim 0.05\,$sr, compared to $\sim 0.6\,$sr for the experimental wavefront. Note that in
both cases the intensity distributions at the focal plane are the same, only the wavefront differs. These results indicate that the transverse wakefield inhomogeneities act as focusing/defocusing electron optics, resulting in the observed complex patterns. A good electron beam quality can be obtained when the transverse inhomogeneities are reduced over the acceleration length, i.e. a few Rayleigh lengths. This suggests that tailoring the wavefront permits the control of the focusing properties of the transverse wakefield. This also explains the results of Ref.~\cite{he13}, where a deformable mirror was used in order to modify the wavefront and to optimize the electron beam quality.


In conclusion, we have observed and explained the complex electron distributions in a high repetition rate laser-plasma accelerator. PIC simulations indicate that these complex distributions are related to distortions of the laser wavefront which cause inhomogeneities in the laser distribution and the transverse wakefield. Our work emphasizes the importance of using realistic laser distributions in simulations in order to understand and reproduce experimental results quantitatively. In addition, while our experiment operates in an extreme regime where the interaction length is large compared to the Rayleigh length, it demonstrates the importance of controlling the laser wavefront over several Rayleigh lengths in laser-plasma interaction experiments. These results might extend to the case of higher energy laser-plasma accelerators, even when self-focusing governs the laser propagation.  

Acknowledgments: This work was funded by the European Research Council under Contract No. 306708, ERC Starting Grant FEMTOELEC and the Agence Nationale pour la Recherche under contract ANR-11-EQPX-005-ATTOLAB


\end{document}